# Investigation on thermal conductivity and viscosity of nanofluids using analytical and machine learning models

Shankar Durgam[1,*] and Ganesh Kadam[1]

[1]Department of Mechanical Engineering, College of Engineering Pune, 411005, India.

*Corresponding author, email: sod.mech@coep.ac.in

## ABSTRACT

Knowledge of thermal properties is essential to design and evaluate thermal systems and processes using nanofluids. This paper presents different analytical models to predict thermal conductivity and viscosity. The efforts have been made to develop machine learning models to predict these properties. An extensive literature survey was carried out to collect thermal properties data of different nanofluids to train and test these machine learning models. The most influential properties like thermal conductivity, diameter, volume concentration of nanoparticles, base fluid thermal conductivity and nanofluid temperature are used as input variables to the thermal conductivity models and molecular weight, diameter and volume fraction nanoparticles, base fluid viscosity and nanofluid temperature are taken as an input variable to the viscosity model. Data is divided into two-part, one part is used to train the models and remaining part is used to test it. Result shows linear regression and ANN model do predict thermal conductivity more closely and ANN predict viscosity more accurately compared to analytical models.

## I. INTRODUCTION

Achieving higher convective heat transfer rate in thermal systems and processes is always been objective. Researchers are worldwide developing alternative techniques to achieve it efficiently. Higher heat transfer rate enables to reduce size, complexity and working cost of the thermal systems and processes. There are two main techniques used to achieve higher convective heat transfer rate viz., Active method and Passive method. In active method external power source is used to achieve higher convective heat transfer rate. Example of active method is use of mechanical aids like use of pump to enhance velocity of heat transfer fluid, surface vibration and jet impingement. In contrast to it passive method need not require such external power source to get desired higher heat transfer rate. Instead of external power source other passive technique like increasing heat transfer area and improving thermal properties of heat transferring fluids are used. Improving thermal properties of a fluid which is used as a heat transfer fluid in a passive technique helps to improve heat transfer rate and it achieved by dispersing nanoparticle mostly metallic origin in a traditional heat transfer fluid. The word nanofluid is firstly coined by Chio and Eastman [1] for a mixture of metallic origin nanoparticles sized less than 100 nm and base fluid. From Several experimental results [2-4] it can be concluded, nanofluids gives higher convective heat transfer rate as a heat transfer fluid than its base fluid**.** Nanofluids promises better thermal properties and could be used as an alternative in automobile, nuclear cooling system, oil and gas and other industries.

Thermal properties of nanofluids like viscosity, density, heat capacity and thermal conductivity are necessary to design and evaluation of thermal systems and processes. Out of these properties' values of density and heat capacity can be calculated using mixture theory. Studies [5-6] have shown this prediction by mixture theory is in acceptable range. Apart from mixture theory different analytical methods are available to calculate viscosity and thermal conductivity but lack of model which well established and could predict viscosity and thermal conductivity much accurately for range of applications. Several authors [7-19] investigated different analytical models to find thermal conductivity and viscosity of nanofluids.

Literature shows that several studies are available to determine thermal conductivity and viscosity of nanofluids using different methods but machine learning models are limited. Therefore the main objective of this investigation is to build up and evaluate machine learning models to predict viscosity and effective thermal conductivity of nanofluids.

## ANALYTICAL MODELS

1. **Analytical models to predict thermal conductivity:**

    i. Hamilton and Crosser Model:
    This model is a modified maxwell model to calculate thermal conductivity by considering nanoparticles shape factor. This Model is applicable to low concentration mixture of solid particles and liquid. In it nanofluids thermal conductivity is contemplate as a function of volume concentration, shape factor of solid particles and thermal conductivity of base fluid.[2]

    $$K_{eff}/k_f = (k_p + 2k_f + 2\phi k_p + 2\phi k_f) / (k_p + 2k_f + 2\phi k_p - 2\phi k_f) \quad (1)$$

    ii. Li and Peterson Model:
    In experimental investigation Li and Peterson found nanofluid thermal conductivity is a function of volume fraction, temperature of mixture and diameter of nanoparticles. It valid for the volume fraction ranging from 0.5 % to 6%. [8] Effective nanofluid thermal conductivity for $Al_2O_3$ based nanofluid is given by Eq. [2].

    $$K_{eff}/kf = 1 + 0.764481\, \phi + 0.01868867\, T - 0.4621417 \quad (2)$$

    The effective thermal conductivity for CuO nanoparticle is given Eq. [3].

    $$K_{eff}/kf = 1 + 3.76108\phi + 0.01792T - 0.30734 \quad (3)$$

    iii. Timofeeva et al. Model:
    Timofeeva et al. Model meant to predict thermal conductivity of liquid and solid shaped nanoparticle mixture [9]. Nanofluids thermal conductivity is considered as function of base fluid thermal conductivity and volume fraction of nanoparticles as given in Eq. [4].

    $$K_{eff} = (1 + 3\phi)k_f \quad (4)$$

iv. Godson et al. Model:
In Godson et al model thermal conductivity of nanofluid is correlated as a linear of volume concentration of nanoparticles and base fluid thermal conductivity [10]. In this study it concluded rise in thermal conductivity is higher at high temperature for same volume fraction as given in Eq. [5].

$$K_{eff} = k_f (0.9692 \phi + 0.9508) \tag{5}$$

v. Minsta et al. Model:
This Minsta et al model is applicable to large temperature range and nanoparticle volume fractions up to 9%. [11]

Minsta et al. established correlation for thermal conductivity of CuO dispersed nanofluids as given in Eq. [6].

$$k_{eff} = k_f (0.99 + 1.74\phi) \tag{6}$$

Correlation of $Al_2O_3$ dispersed nanofluid is given in Eq. [7].

$$k_{eff} = k_f (1 + 1.72 \phi) \tag{7}$$

2. **Analytical model to predict viscosity:**
   i. Einstein Model:
   Einstein is a pioneer of solid and liquid properties prediction [12]. In this model viscosity of mixture (solid and liquid) is considered as a function of volume concentration of the solid particles and base fluid viscosity. Though this model is not designed to calculate viscosity of nanofluids still it gives good result at a low volume concentration up to 0.02 of the nanoparticles. In this model solid particles are considered as a rigid and chargeless. Einstein model viscosity is given by Eq. [8].

$$\mu = \mu_0 (1 + 2.5 V) \tag{8}$$

   ii. Brinkman model:
   Brinkman is an improved Einstein model which taking an account of solid and liquid particles interaction. In this model like Einstein model viscosity of mixture is considered function of solid particles volume concentration and base fluid viscosity. Unlike Einstein model this model can predict viscosity for a range of concentration [13]. Brinkman model viscosity is given by Eq. [9].

$$\mu = \frac{\mu o}{(1-\phi)^2} \tag{9}$$

   iii. Krieger and Dougherty model:

This model does consider interaction between liquid and solid particles and assumes solid particle rotate about its own mass center. Viscosity is considered as a function of volume concentration of the solid particles and viscosity of base fluid [14]. Krieger and Dougherty model viscosity is given by Eq. [10].

$$\mu = \mu_0 \left(1 - \frac{\phi}{0.5}\right)^{-1.25} \tag{10}$$

iv. G. K. Batchelor model:

Batchelor model considers Brownian motion effect on viscosity of mixture. Viscosity is considered as a function of volume concentration of the solid particles and viscosity of base fluid [15]. The viscosity by Bachelor model is given by Eq. [11].

$$\mu = \mu_0 (1 + 2.5\phi + 6.2\phi^2) \tag{11}$$

v. Ward model:

Ward viscosity model equation is developed by equation fitting with the experimental data and it suitable to predict viscosity up to the 35 % volume concentration of nanoparticles. It is more suitable to predict viscosity of two phase nanofluids [16]. The viscosity by Ward model is given by Eq. [12].

$$\mu = \mu_0 [1 + (2.5\phi) + (2.5\phi)^2 + (2.5\phi)^3 + (2.5\phi)^4 + ...] \tag{12}$$

vi. Tseng and Chen model:

Like previous analytical model here Viscosity is considered as a function of base fluid viscosity and volume concentration of solid particles [13]. The viscosity by Tseng and Chen model is given by Eq. [13].

$$\mu = \mu_0 (0.4513 e^{0.6965\phi}) \tag{13}$$

vii. Chandrasekar et al model:

This model is developed from Maxwell and Einstein model without restoring it. Values of b and n is calculated by fitting experimental data and its value is 5300 and 2.8 respectively. Like previous analytical model in it Viscosity is considered as a function of volume concentration of solid particles and base fluid viscosity [14]. The viscosity by Chandrasekar et al. model is given by Eq. [14].

$$\mu = \mu_0 \left(1 + b \left(\frac{\phi}{1-\phi}\right)^n\right) \tag{14}$$

 viii. Ghasemi and Karimipour model:
  ix. Ghasemi and Karimipour considered nanofluid viscosity as a function of volume fraction of the nanoparticles, temperature of mixture and viscosity of the base fluid [19]. The viscosity by Ghasemi and Karimipour model is given by Eq. [15].

$$\mu = \mu_0(-1.735T^{-0..17} - 0.027\phi^{0.418} + 0.039\phi^{1.543}T^{-0.033} + 2.956) \tag{15}$$

## II. Machine learning Model Methodology

Most of the analytical methods used to predict viscosity and thermal conductivity of a nanofluids gives accurate result for specific range of temperature and volume fraction of the solid nanoparticles. Even some are meant for specific type of nanofluids. Nanofluid promises best alternative to conventional fluid as a heat transfer fluid still, 25-years of research has not yet yielded into analytical model which is well established and valid for all type of nanofluids to predict viscosity and thermal conductivity. Main reason behind this could be the lack of well-defined relationship between input output variables and consideration of main attributers to rise in thermal conductivity like nanoparticle dispersion, Brownian motion and thermophoresis. Few analytical models considered effect of Brownian motion for predicting thermal conductivity but has ignored other effect like nanoparticle dispersion and thermophoresis. Agglomeration of nanoparticle is another obstacle in both use of nanofluid and predicting its thermal properties. Though few analytical models have considered effect of agglomeration of nanoparticle but this phenomenon is dynamic which makes model difficult to use in particle cases.

Machine learning models can be used as predictive model in case correlation between output and input variables is not well defined. The characteristics of machine learning models to predict output even in case correlation between output and input variables not well-defined sets as an ideal candidate to predict viscosity and thermal conductivity of nanofluids. Machine learning models mainly focus on predictive modeling by minimizing the error of a model or making most accurate prediction possible, at expense of expansibility. Advantage of machine learning solutions is its ability of solving complex problem, easy applicable and ability to process large amount of data. In recent past focus of researchers has been shifted to machine learning models as alternative method to predict thermal properties.

Fig. 1 shows typical representation of machine learning model. It consists of mainly three steps.

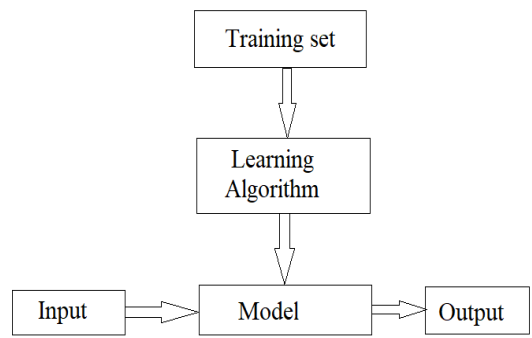

Fig.1 Flow chart of machine learning model

1. Collection of data:
   First step in developing of machine learning model is collection of data as dependent variable and respective independent variable. Result of models are mainly dependent on the quality of the data so to get desirable result data must be collected from well verified sources.
2. Feeding data to the learning algorithm:
   Once data collected as input variable and output variable it divided into two part as testing and training data set. Training data set is given as an input to learning algorithm in which initially weight assigned randomly. With the help of training data final weight are adjusted by using backpropagation.
3. Testing of model:
   Once weights in model calculated by using training data set model can tested using testing data set. The accuracy of model is depending on the accuracy of training data set.

In this analysis two machine learning models Linear regression and Artificial neural network (ANN) used to predict viscosity and effective thermal conductivity of nanofluids.

1. Linear regression:
   Linear regression model considerers linear correlation between output and input variables. The linear regression model to predict thermal conductivity is given by Eq. [16] and the model to predict viscosity of nanofluid is given by Eq. [17].

$$K_{eff} = f(d, k_f, k_p, T, \phi) \tag{16}$$

$$\mu_0 = f(d, \mu_0, M, T, \phi) \tag{17}$$

Outcomes of the model is dependents on how close values of weight in model to the ideal value. here ideal model considered model which gives prediction close to the experimental value assuming input and output variables has a linear correlation. The value of weight is calculated by solving normal equation. Here X is input variables, θ represent weights and Y is output, as given in Eq. [18].

$$\Theta = (X^T X)^{-1} X^T Y \tag{18}$$

2. ANN model:

Artificial Neural Network (ANN) first time introduced in 1940 by McCulloh and Pitts [20]. They got popularity as researchers started to use backpropagation technique to find out weights in the network. ANN model is inspired from processing of the human brain and find out pattern within the data. ANN is a set of the neurons arranged in pattern like a biological neuron which meant to process data and weights are equivalent to synapses which interconnects neurons and approximates activity in some of the human brain. ANN has ability to learn patterns from the training data set.

ANN is able to establish nonlinear multidimensional relationship in input and output variables. Unlike linear regression, ANN model can predict viscosity and thermal conductivity even input and output variables has a nonlinear, multidimensional relationship. Fig.2 shows general Architecture of ANN.

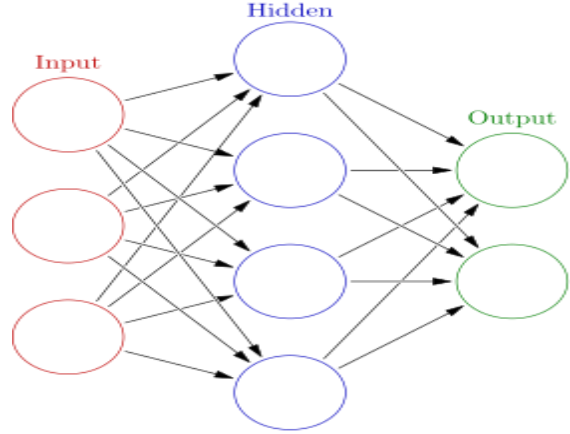

Fig.2 ANN Architecture

Typical ANN has three layer first is Input layer second Hidden layer and last is Output layer. In This study ANN with five input neurons which act as an input variable is used. In thermal conductivity model volume concentration, diameter of the nanoparticles, base fluid and nanoparticles thermal conductivity, temperature of mixture are used as input variable. For viscosity model base fluid viscosity, diameter, volume concentration, molecular weight of nanoparticles and temperature of the mixture is used as input variables. Values of each input and output variable in neural network is kept between 0 to 1. Numerical value of this input variables is act as input layer in ANN. In this study model with two layers in hidden layer and 12 neurons in each hidden layer is used. The output of output layer nodes in the predicted output which is viscosity and effective thermal conductivity of the nanofluids. The output of each layer is going through an activation function. Activation function mainly used to model nonlinear behavior. Sigmoid function is used as an activation function and it is defined as given by Eq. [19].

$$f(Z_i) = 1 \backslash (1+e^{Z}) \tag{19}$$

The neural network is trained with training data set by using backpropagation method. The term backpropagation referred to calculate error at output layer and it propagated backward to adjust weights in neural network.

Numerical value of each neuron in neural network is defined as given by Eq. [20].

$$Z_{ij} = \Theta_{ij} X_{ij} \quad [20].$$

Here Z represent value of neuron in $j^{th}$ neuron in $i^{th}$ layer, Θ represents weight, and X is input value form previous neurons or input variable.

## III.  Experimental Data

Extensive literature survey conducted to collect viscosity data set and thermal conductivity data set of nanofluids. In this particular study of thermal conductivity data from 473 experimental results for six different types of nanofluids such as, $Al_2O_3$-EG, $Al_2O_3$-$H_2O$, $TiO_2$-H2O, CuO-$H_2O$, CuO-EG, $TiO_2$-EG collected from different research works [21-31]. Also, viscosity data set of 443 experimental result of six different types of nanofluids like $Al_2O_3$-$H_2O$, $Al_2O_3$-EG, CuO-$H_2O$, CuO-EG, $TiO_2$-EG, $TiO_2$-$H_2O$ collected from different research works [32-44].

Total data of both viscosity and thermal conductivity is divided into data set which is training and testing data set. Out of total data 70% data used to train the model and remaining 30% used for testing. Training data set is given as a input to learning algorithm to find out relationship between output and input variables by adjusting weight using backpropagation technique. To minimize possibility of low convergence rates numerical values of output and input normalized in range of 0 to 1.

## IV. Results and discussion

**Performance assessment indices:**

Different plots comparing viscosity and thermal conductivity outcomes by experimental result, different analytical models and machine learning models are shown. Other than that statistical quality measure tool relative root mean square (RRMSE) is calculated. Model considered excellent if RRMSE<10 %, Fair if 20% < RRMSE<30%, good if 10% < RRMSE <20%, and poor if RRMSE >39% [17].

**Thermal Conductivity model Result:**

The predicted thermal conductivity of different nanofluids by machine learning and the analytical models are compared with experiment result in Fig.3 to Fig 5. Fig.3 is a comparison plot of $Al_2O_3$-EG nanofluid thermal conductivity by analytical models of Li and Peterson, Timofeeva et al. [Fig. 3 (a)] and Godson et al and Minsta et al. {Fig. 3 (b)], machine learning models and experimental result. Testing point for this result is selected from testing data set by keeping diameter of nanoparticles as 50 nm and volume fraction as 0.02. Plots clearly shows both machine learning models viz., (i) Linear regression and (ii) ANN predict thermal conductivity close to the experimental value compared to prediction by the analytical models. It is noticed that the ANN and linear regression thermal conductivity values are well match with the experimental results from 24 to 35 °C. However, after the temperature of above 35 °C the experimental thermal conductivity values are not very close to the predicted analytical and machine learning model values.

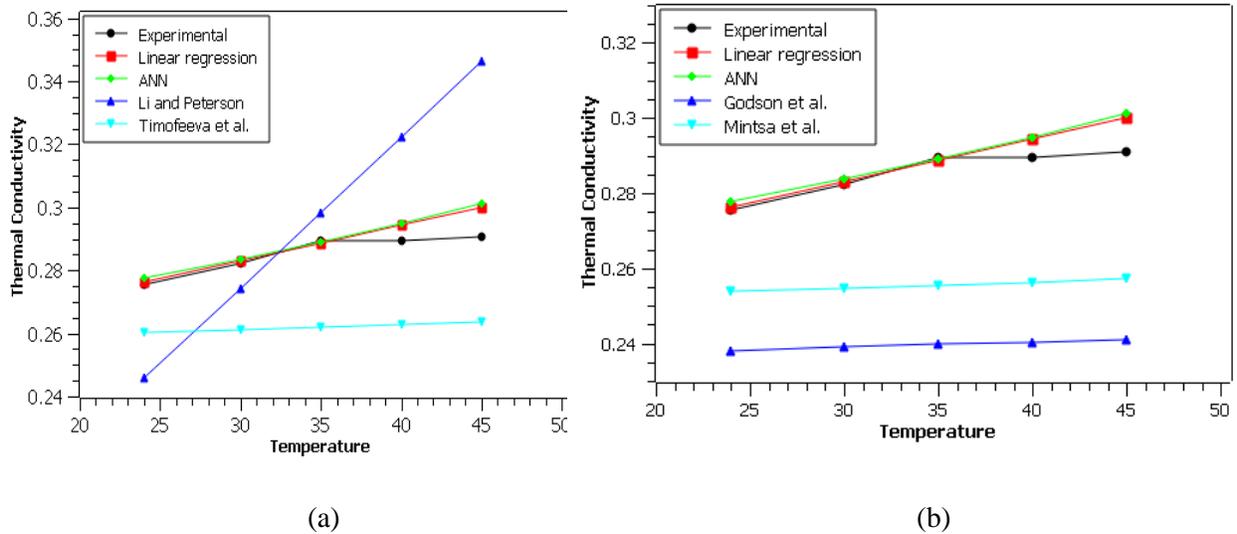

(a)                                                                                                         (b)

Fig.3 Comparison of effective thermal conductivity predicted by analytical model and machine learning model with experimental result of $Al_2O_3$ - EG nanofluids.

Fig. 4 shows a comparison plot drawn between thermal conductivity and volume fraction values for CuO-Water nanofluid by analytical models, machine learning models and experimental result. Fig. 4 (a) shows a comparison of thermal conductivity in analytical models of Hamilton and Crosser, and Timofeeva et al., machine learning and experimental results. Fig. 4 (b) shows a comparison of thermal conductivity in analytical models of Godson et al., and Minsta et al., machine learning and experimental results. Testing point for this plot are selected from testing data set by keeping diameter of nanoparticles as 22 nm and temperature as 29 °C. Result shows prediction of thermal conductivity values by both machine learning model and analytical models diverge from experimental result but RRMSE value (Fig.6) is lower than 10 for all the models which makes all models are within excellent modes to predict thermal conductivity of nanofluids [17]. Moreover the divergence of thermal conductivity between analytical model, machine learning and experimental values in case of CuO – Water are marginal of the order of 0.1

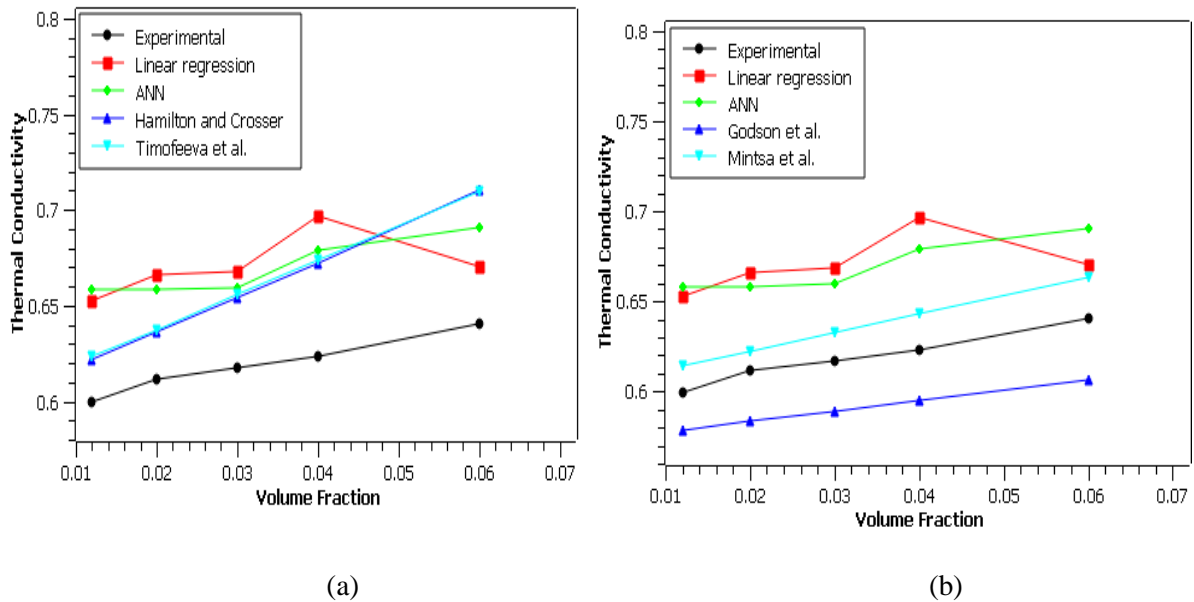

(a) (b)

Fig.4 Comparison of effective thermal conductivity predicted by analytical model and machine learning model with experimental result of CuO-Water nanofluids.

Fig.5 is a comparison plot of $TiO_2$-Water nanofluid thermal conductivity predicted by analytical models of Li and Peterson, and Timofeeva et. al. [Fig. 5(a)], machine learning models and experimental results. Whereas, Fig. 5(b) shows analytical models of Godson et al., and Minsta et. al., machine learning models and experimental results. Testing point for this plot are selected from testing data set by keeping diameter of nanoparticles as 25 nm and temperature as 25 °C. Plots clearly shows both machine learning models Linear regression and ANN predict thermal conductivity close to the experimental value compared to prediction by the analytical models. Hence the prediction of thermal conductivity of nanofluids such as $TiO_2$-Water by machine learning models viz., linear regression and ANN are preferred over analytical models.

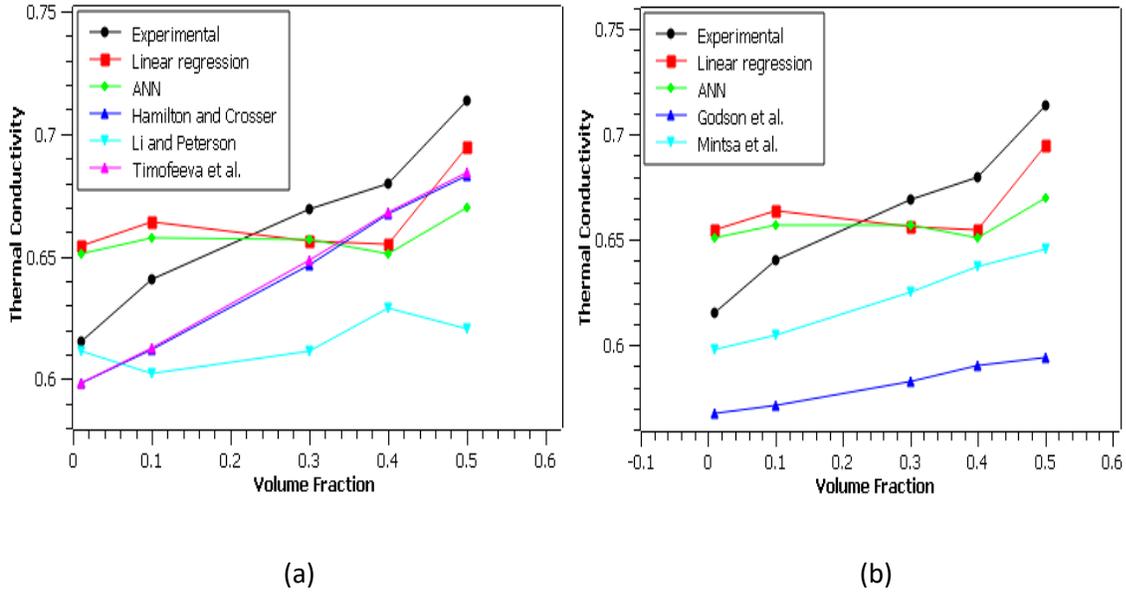

(a)            (b)

Fig.5 Comparison of effective thermal conductivity predicted by analytical model and machine learning model with experimental result of $TiO_2$-Water nanofluid

Fig. 6 show the RRMSE values of analytical and machine learning model that is compared for $Al_2O_3$-EG, CuO-Water and $TiO_2$-Water nanofluids. Result clearly shows, for all three type of nanofluids values of RRMSE for both linear regression and Artificial neural network model is lower than 10 which makes them excellent model to predict thermal conductivity [17]. The analytical models do predict thermal conductivity with good accuracy for CuO water nanofluids but not for other two type of nanofluids.

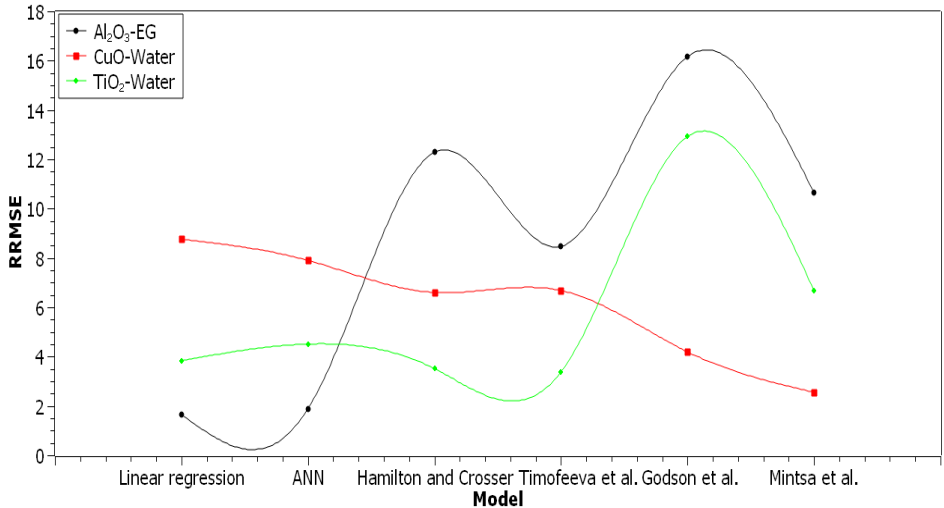

Fig.6 Comparison of RRMSE values of analytical and machine learning model

**Viscosity model Result:**

The prediction of viscosity of different nanofluids by machine learning model and Analytical models is compared with experiment result in Fig.7 to Fig 9. Fig. 7(a) shows comparison plot of $Al_2O_3$-EG nanofluid viscosity by analytical models of Einstein, Bachelor, Brinkman and Ward et al. and machine learning models with experimental result. Whereas, Fig. 7(b) shows comparison plot of $Al_2O_3$-EG nanofluid viscosity by analytical models of Krieger et al., Tseng et al., Chandrasekar et al. and Ghasemi et al. and machine learning models with experimental result. Testing point for this plots are selected from testing data set by keeping diameter of nanoparticles as 13 nm and temperature as 25 °C. Plots clearly shows only ANN model predict viscosity close to the experimental value compared to prediction by the analytical and linear regression model for a temperature range of 20 – 80 °C. Hence for a nanoparticle of 13 nm diameter and the temperature of 25 °C ANN model is suitable compared with all analytical models and regression model considered in the present study.

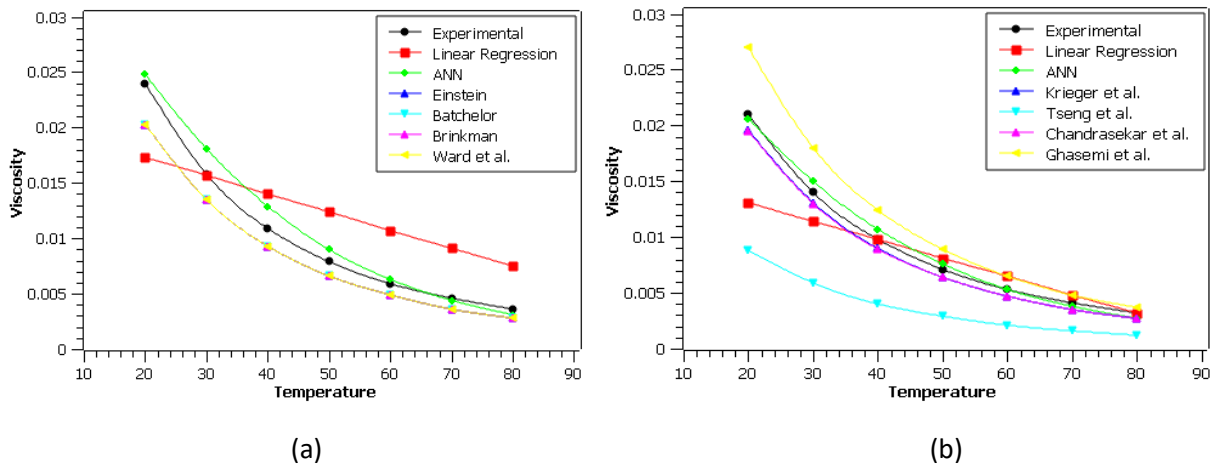

(a) (b)

Fig.7 Comparison of viscosity prediction by analytical model, machine learning model and experimental result of $Al_2O_3$-EG nanofluids.

Fig. 8 (a) is a comparison plot of CuO-EG nanofluid viscosity by analytical models of Einstein, Bachelor, Brinkman and Ward et al., and machine learning models with experimental result. Fig. 8 (b) is a comparison plot of CuO-EG nanofluid viscosity by analytical models of Krieger et al., Tseng et al., Chandrasekar et al. and Ghasemi et al. and machine learning models with experimental result. Testing point for this plot are selected from testing data set by keeping nanoparticle diameter 50 nm and volume fraction 0.015. It is noticed from these plots that only ANN model predict viscosity close to the experimental value compared to prediction by the analytical and linear regression model for a given temperature range of 20 – 80 °C.

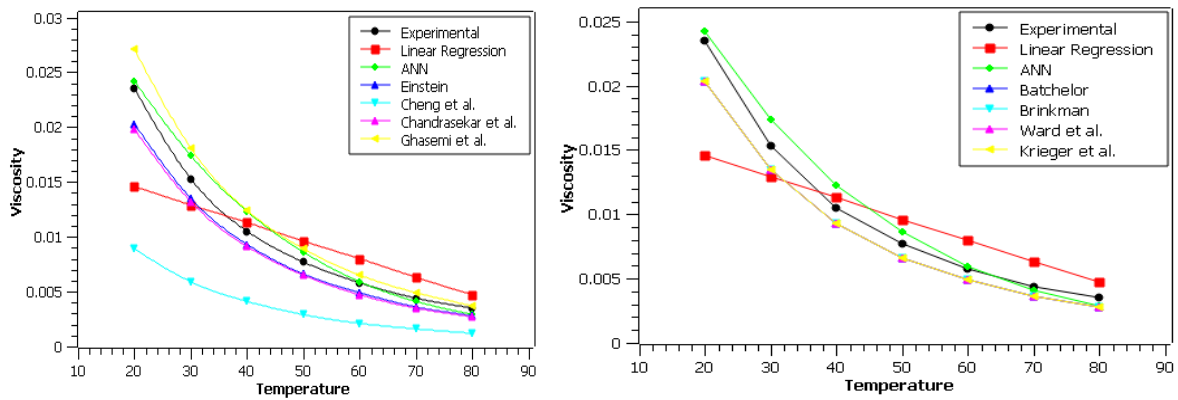

(a) (b)

Fig.8 Comparison of viscosity prediction by analytical model and machine learning model with experimental result of CuO-EG nanofluids.

Fig. 9 (a) shows comparison plot of $CeO_2$-EG nanofluid viscosity prediction by analytical models of Einstein, Bachelor, Brinkman and Ward et al., and machine learning models with experimental result. Similarly, Fig. 9 (b) is a comparison plot of $CeO_2$-EG nanofluid viscosity by analytical models of Krieger et al., Tseng et al., Chandrasekar et al. and Ghasemi et al. and machine learning models with experimental result. Testing point for this plot are selected from testing data set by keeping diameter of nanoparticles as 50 nm and volume fraction as 0.015. It is found from these plots that only ANN model predict viscosity close to the experimental value compared to prediction by analytical and linear regression model for a given range of temperature.

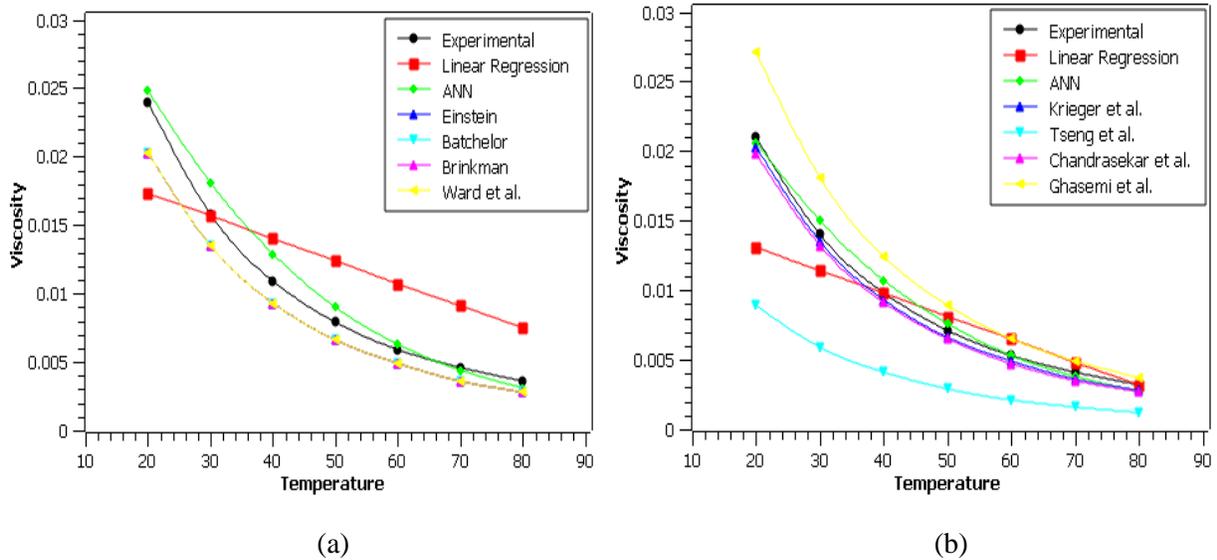

(a) (b)

Fig.9 Comparison of viscosity prediction by analytical model, machine learning model and experimental result of $CeO_2$-EG nanofluids.

Fig. 10 shows a RRMSE values of analytical and machine learning model compared for $Al_2O_3$-EG, CuO-EG and $CeO_2$-EG nanofluids. Results indicates that, for all three nanofluids studied in the present work the value of RRMSE is lowest in case of artificial neural network model which makes it an excellent model to predict viscosity of the nanofluids.

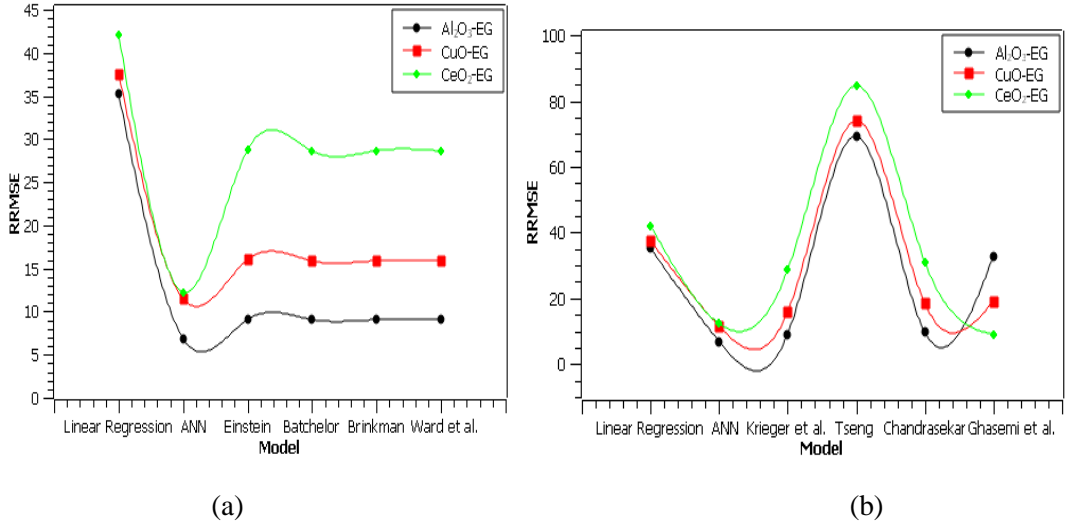

(a)                         (b)

Fig.10 Comparison of RRMSE values of analytical and machine learning models

## V. Conclusion

In this study machine learning model is build up to predict numerical values of viscosity and thermal conductivity of nanofluids using experimental data from literature. The influential properties like thermal conductivity nanoparticles and base fluid, volume concentration and diameter of nanoparticle, temperature of nanofluid are used as an input variable in thermal conductivity model. In viscosity model influential properties like base fluid viscosity, diameter, volume concentration and molecular weight of nanoparticles and temperature of nanofluid are considered input variables in the viscosity models.

Statistical quality measure relative root means square error and Different plots comparing prediction of viscosity and thermal conductivity by analytical and machine learning model is used to evaluate all models. By comparing result of thermal conductivity prediction and RRMSE values it is found that linear regression and machine learning models predict thermal conductivity much accurately compared to prediction by analytical models. In case of viscosity prediction RRMSE value is lowest for artificial neural network compared to analytical and linear regression model.

As an alternative modeling approach to predict viscosity and effective thermal conductivity of the nanofluids, results indicate that ANN could be useful to reduce the cost, time and nanofluids incurred using conventional experimental method

**Nomenclature:**
$K_{eff}$ = nanofluids effective thermal conductivity
$T$ = Temperature
$K_f$ = Thermal conductivity of the base fluid
$\mu$ = Viscosity of the nanofluids
$d$ = Nanoparticle diameter
$\mu_0$ = base fluid viscosity
$\phi$ = Nanoparticles volume fraction

$K_p$ = Nanoparticle thermal conductivity

M = molecular weight of nanoparticles

**REFERENCES:**

[1] S. U. Choi, J. A. Eastman, Enhancing thermal conductivity of fluids with nanoparticles, Tech. rep., Argonne National Lab., IL (United States) (1995).

[2] Jaafar Albadr, Satinder Tayal, Mushtaq Alasad, Heat transfer through heat exchanger using Al2O3nanofluidat different concentrations, case studies in thermal engineering Volume 1, Issue 1, October 2013, Pages 38-44.

[3] Poppy Puspitasari1,2, Avita Ayu Permanasari1 , Maizatul Shima Shaharun3 , and Dewi Izzatus Tsamroh1, Heat Transfer Characteristics of NiO Nanofluid in Heat Exchanger, AIP Conference Proceedings 2228, 030024 (2020).

[4] B. Jefferson, Raja Bose1, L. Godson Asirvatham2 and Michael N Kumar, Experimental Convective Heat Transfer Studies on Graphene Nanofluid for the Cooling of Next Generation Electronic Components, International Journal of Applied Engineering Research ISSN 0973-4562 Volume 12, Number 19 (2017) pp. 8534-8539

[5] Tun-Ping Teng and Yi-HsuanHung, Estimation and experimental study of the density and specific heat for alumina nanofluid, Journal of Experimental Nanoscience, 2014 Vol. 9 no 7 707 718

[6] R. Hamilton, O. Crosser, Ice fundamentals, Int Heat Mass Trans2(1962) 187–189

[7] C. H. Li, G. Peterson, Experimental investigation of temperature and volume fraction variations on the effective thermal conductivity of nanoparticle suspensions, Journal of Applied Physics 99 (8) (2006)084314

[8] Timofeeva, A. N. Gavrilov, J. M. McCloskey, Y. V. Tolmachev, S. Sprunt, L. M. Lopatina, J. V. Selinger, Thermal conductivity and particle agglomeration in alumina nanofluids: experiment and theory, Physical Review E 76 (6) (2007) 061203

[9] L. Godson, B. Raja, D. Mohan Lal & S. Wongwises (2010) Experimental Investigation on the Viscosity and thermal conductivity of Silver-Deionized Water Nanofluid, Experimental Heat Transfer, 23:4, 317-332, DOI: 10.1080/08916150903564796

[10] Minsta HA, ROY G, Cong TN, Doucet D, New temperature dependant thermal conductivity data for water based nanofluids. Int J Therm Sci 48:363-371

[11] Einstein A. Eine neue Bestimmung der Moleküldimensionen. Ann. Phys. 1906; 324:289-306

[12] By G. K. BATCHELOR, The effect of Brownian motion on the bulk stress in a suspension of spherical particles, J. Fluid Mech. (1977), vol. 83, part 1, pp. 97-117

[13] Brinkman, H. C. (1952): The viscosity of concentrated suspensions and solution. Journal Chemical Physics, vol. 20, pp. 571-581

[14] Krieger, I. M. and Dougherty, T. J. 1959. A mechanism to non-Newtonian flow in suspensions of rigid spheres. Trans. Soc. Rheology, 3, pp. 137-152.

[15] N.-S. Cheng, A.W.-K. Law, Exponential formula for computing effective viscosity, Powder Technol. 129 (2003) 156–160

[16] Chandrasekar, M., Suresh, S. and Bose, A.C., "Experimental investigations and theoretical determination of viscosity and thermal conductivity of Al2O3/water nanofluid", Experimental Thermal and Fluid Science, Vol. 34, No. 2, (2010), 210-216

[17] Ghasemi S, Karimipour A. Experimental investigation of the effects of temperature and mass fraction on the dynamic viscosity of CuO-paraffin nanofluid. Appl Therm Eng. 2018; 138:189-197.

[18] W.S. McCulloh, W. Pitts, A logical calculus of the ideas immanent in nervous activity, Bull. Math. Biophys. 5 (1943)

[19] A. M. Hussein, R. Bakar, K. Kadirgama, K. Sharma, Experimental measurement of nanofluids thermal properties, International Journal of AutomotiveandMechanicalEngineering7(2013)850.

[20] C. H. Li, W. Williams, J. Buongiorno, L.-W. Hu, G. Peterson, Transient and steady-state experimental comparison study of effective thermal conductivity of al2o3/ water nanofluids, Journal of heat Transfer 130 (4) (2008).

[21] K. Mahanpour, S. Sarli, M. Saghi, B. Asadi, R. Aghayari, H. Maddah, Investigation on physical properties of al2o3/water nano fluid, Journal of Materials Science & Surface Engineering, http://www.jmsse. org (2015).

[22] J. Buongiorno, D. C. Venerus, N. Prabhat, T. McKrell, J. Townsend, R. Christianson, Y. V. Tolmachev, P. Keblinski, L.-w. Hu, J. L. Alvarado, et al., A benchmark study on the thermal conductivity of nanofluids, Journal of Applied Physics 106 (9) (2009) 094312.

[23] S. Murshed, K. Leong, C. Yang, Investigations of viscosity and thermal conductivity of nanofluids, International journal of thermal sciences 47 (5) (2008)560–568.

[24] I. Fernández, R. Valiente, F. Ortiz, C. J. Renedo, A. Ortiz, Effectoftio and zno, nanoparticles on the performance of dielectric nanofluids based on vegetable esters during their aging, Nanomaterials 10 (4) (2020)692.

[25] M. H. Esfe, The investigation of effects of temperature and nanoparticles volume fraction on the viscosity of copper oxide-ethylene glycol nanofluids, Periodical Polytechnical Chemical Engineering 62 (1) (2018) 43–50.

[26] L. Fedele, L. Colla, S. Bobbo, Viscosity and thermal conductivity measurements of water-based nanofluids containing titanium oxide nanoparticles, International journal of refrigeration 35 (5) (2012) 1359–1366.

[27] M. H. Esfe, A. Karimipour, W. M. Yan, M. A. kbari, M. R. Safaei, M. Dahari, Experimental study on thermal conductivity of ethylene glycol based nanofluids containing al2o3 nanoparticles, International Journal of Heat and Mass Transfer 88 (2015) 728–734.

[28] X. Wang, X. Xu, S. U. Choi, Thermal conductivity of nanoparticle fluid mixture, Journal of thermophysics and heat transfer 13 (4) (1999) 474–

480.

[29] A. Zendehboudi, R. Saidur, A reliable model to estimate the effective thermal conductivity of nanofluids, Heat and Mass Transfer55(2) (2019) 397–411.

[30] M. Mehrabi, M. Sharifpur, J.P. Meyer, Viscosity of nanofluids based on a machine learning model, International Communications in Heat and Mass Transfer 43 (2013) 16–21

[31] Y. Raja Sekhar & K.V. Sharma, Study of viscosity and specific heat capacity characteristics of water-based Al2O3 nanofluids at low particle concentrations, Journal of Experimental Nanoscience, 2015Vol. 10, No. 2, 86–102

[32] Houda Jalali1, and Hassan Abbassi, Analysis of the Influence of Viscosity and Thermal Conductivity on Heat Transfer by Al2O3-Water Nanofluid, FDMP, vol.15, no.3, pp.253-270, 2019

[33] A. Prakash1, S. Satsangi1, S. Mittal1, B Nigam1, P. K. Mahto1, Bibhu P Swain, Investigation on Al2O3 Nanoparticles for Nanofluid Applications- A Review, IOP Conf. Series: Materials Science and Engineering 377 (2018) 012175

[34] Kamaldeep Singh, Sumeet Sharma, Experimental Study of Thermophysical Properties of Al2O3Water Nanofluid, IJRMET Vol. 3, Issue 2

[35] Minsuk Kong1 and Seungro Lee, Performance evaluation of Al2O3nanofluid as an enhanced heat transfer fluid, Advances in Mechanical Engineering2020, Vol. 12(8) 1–13

[36] Eda Feyza AKYÜREK, Kadir GELİŞ2and Bayram ŞAHİN, The Experimental Investigation of The Viscosity of The Al2O3 Nanofluid, Eastern Anatolian Journal of Science Volume V, Issue II, 2019, 26-35

[37] Shre K. Menakshi, Pradep E. JayaSudhan, Development and characterization of aluminum oxide nanofluids as coolant s for heat transfer plications, ChemXpress 9(3), 266-272, (2016)

[38] Hajir Karimi1, FakheriYousefi, Mahmood Reza Rahimi, Correlation of Viscosity in Nanofluids using Genetic Algorithm-neural Network (GA-NN), International Journal of Chemical and Molecular EngineeringVol:5, No:1, 2011

[39] Sandeep Kumar, Gurpreet Singh Sokhal and Jaspreet Singh, Effect Of Cuo-Distilled Water Based Nanofluids On Heat Transfer Characteristics And Pressure Drop Characteristics., ISSN: 2248-9622, Vol. 4, Issue 9

[40] k. Shree Meenakshi, E. Pradeep, Jaya Sudhan, Preparation and Characterization of Copper Oxide -Water Based Nanofluids by One Step Method for Heat Transfer Applications, Chemical Science Transactions2015, 4(1), 127-132

[41] Karunakar Singh, Shiv Kumar, An Experimental Study on Characterization of CuO/Water Nanofluid, IRJET Volume: 07 Issue: 08 (400-404)

[42] Sarah Simpson, Austin Selfhood, Chris Golden and Saeid Vafaei, Nanofluid Thermal Conductivity and Effective Parameters, Appl. Sci. 2019, 9, 87